# Exploring the Correlation between Urban Microclimate Simulation and Urban Morphology: A Case Study in Yeongdeungpo-gu, Seoul

Yan Xiang[1], Danni Chang[1], Jieli Cheng[2]
[1]School of Design, Shanghai Jiao Tong University, Shanghai, China
[2]Graduate School of Environmental Studies, Seoul National University, Seoul, South Korea
yanxiang@sjtu.edu.cn (Yan Xiang), dchang1@sjtu.edu.cn (Danni Chang)

*Abstract* - **Different social backgrounds and planning policies give rise to diverse urban morphologies. These morphologies influence urban microclimate factors and contribute to the formation of unique local microclimates, particularly in terms of outdoor temperature. In recent times, the heat island effect has gained increasing significance during the summer season. Therefore, this study aims to explore the correlation between urban microclimate simulation and urban morphology within the context of the heat island effect. Specifically, we investigate how the outside temperature varies across different types of residential buildings in Yeongdeungpo-gu, Seoul, South Korea, during the summer period. We compare temperature conditions using a multi-dimensional system of building clusters' morphological indices and employ ENVI-met software for simulation purposes. The results of the urban microclimate simulation are comprehensively analyzed, revealing a significant finding: high-rise residential buildings exhibit considerably higher outdoor temperatures compared to low-rise residential buildings. Furthermore, the presence of open spaces plays a crucial role in mitigating high neighborhood temperatures. By deriving insights from these findings, we aim to provide valuable conclusions to support city managers in making informed decisions.**

## I. INTRODUCTION

With the rapid development of cities and their constant evolution, coupled with the ever-increasing influx of data, urban morphology is becoming increasingly complex. In this context, urban computing, particularly in the realm of urban big data analysis and insights into human behavior, has gained significant importance. In recent years, the proliferation of high-density and high-volume-rate built-up areas has significantly altered the underlying surface of cities. Combined with substantial anthropogenic heat emissions, this has resulted in a noticeable 'urban heat island effect' between urban areas and their suburbs [1]. This effect has far-reaching consequences for the urban environment, human settlements, and physical and mental health. While numerous studies have explored the impacts of increasing urban building density on urban microclimate, particularly focusing on the significant heat island effect [2], few have delved into the influence of urban morphology on outdoor temperature, which is also a contributing factor to the urban heat island effect [3].

Therefore, our study aims to investigate the relationship between outdoor air temperature, urban microclimate, and various urban forms using ENVI-met, a 3D modeling software that enables high-resolution simulation of outdoor microclimates. This software has become a fundamental tool in urban and building design research. Through this process, we examine a specific residential area in Seoul and analyze how outdoor temperatures differ between two types of residential buildings during the summer. Our findings reveal that morphological indicators at different levels exert a synergistic effect on the microclimatic impact of street space morphology. This comparative analysis of microclimatic characteristics at a medium spatial scale provides valuable insights for optimizing urban spatial morphology in future urbanization development, enhancing urban climate adaptability.

In conclusion, our study integrates the realms of data-driven analysis, urban morphology, and innovative computing tools to shed light on the intricate interplay between urban form, microclimates, and outdoor temperature. It underscores the imperative of informed, morphology-conscious urban planning and design in the face of rapid urbanization, ultimately charting a course towards more livable, climate-resilient cities.

## II. RELATED WORK

In recent years, the impact of accelerated urbanization on urban microclimate has become a significant area of research. Urban areas exhibit diverse morphologies and densities, which in turn influence the thermal and radiative properties of their surfaces. As a result, microclimate conditions within cities can vary considerably [4]. One notable consequence of these variations is the occurrence of what is known as the "urban heat island" effect, where cities experience higher air and surface temperatures compared to surrounding rural areas [5].

The urban heat island effect, a well-documented feature of urban climates, has been the subject of extensive scholarly investigation since the early 19th century. Scholars have made substantial contributions in delineating the characteristics and underlying causes of this phenomenon. It has been firmly established that the urban heat island effect is intricately linked to factors such as urban anthropogenic heat emissions, the specific nature of ground surfaces, urban structural design, vegetation distribution, and population density. As urbanization continues its relentless expansion, both the intensity and geographical extent of the urban heat island effect continue to amplify [8].

Additionally, the presence of tall and dense buildings in urban areas poses challenges to air circulation, impeding the dispersion and settling of gaseous and particulate pollutants. This leads to the exacerbation of atmospheric pollution, particularly in the form of particulate matter (PM) pollutants [9]. The intensification of these urban problems emphasizes the growing importance of studying urban ecology and finding sustainable solutions. It becomes increasingly evident that many environmental issues in cities stem from the inadequate design of urban morphology and industrial layouts, resulting in an unbalanced spatial distribution of urban resources and energy [10].

In light of these challenges, it becomes crucial to incorporate the impact of spatial form on urban microclimate when designing cities. Considering the interplay between urban morphology and microclimate can help create more livable and sustainable urban environments. In recent years, there have been remarkable advancements in computational resources, enabling simulation-based computational methods to emerge as primary research tools for urban microclimate studies [11]. These methods offer the ability to simulate and analyze urban microclimates in high resolution, providing valuable insights for urban planning and design decisions that enhance urban climate adaptability and sustainability.

## III. METHODOLOGY

This study focuses the outdoor temperature of different residential blocks in Seoul, South Korea, as the research subjects, summarizes the research status and trend of urban microclimate and outdoor temperature related fields, and uses the ENVI-met software as numerical simulation method and application as shown in Figure 1.

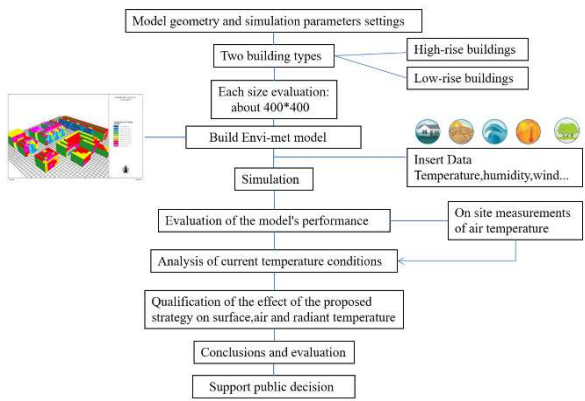

Fig. 1. Framework of research and simulation methodology.

To address the issue of urban heat island, this study begins by establishing a 3D digital surface model of buildings, which serves as fundamental data. Focusing on select areas within Seoul's Yeongdeungpo district, satellite remote sensing images and CAD models are utilized for interpretation and geometric correction. Information such as building area and geographic coordinates is determined, followed by collecting data on building heights for different residential structures. This forms the basis for constructing a comprehensive digital surface model database. Subsequently, the research involves partitioning the area based on natural properties such as streets and other features, enabling the differentiation of various structural shapes. The classification research then focuses on high-rise and low-rise buildings.

*1) Selection and Classification Principles for the Research Site:* The research site is chosen to encompass a variety of building types, including high-rise and low-rise structures. This selection aims to eliminate interference factors stemming from different land masses, such as varying building densities, traffic conditions, wind speeds, and outdoor temperatures. When classifying building types, plots with well-defined boundaries are preferred. Considering these factors, the Yongdeungpo Area in Seoul is chosen as the research site. Surrounding conditions and satellite maps illustrating the area are presented in Figure 2 and 3, respectively.

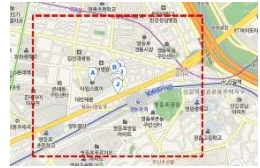 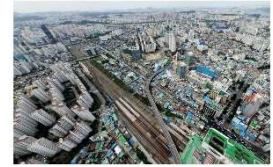

Fig. 2. Surrounding condition map.  Fig. 3. Satellite map.

By assessing the overall height of building cluster, using road and the nature building cluster characteristic as as a natural boundary to classify. Two residential building types(a. high-rise building and b. low-rise building)are selected.

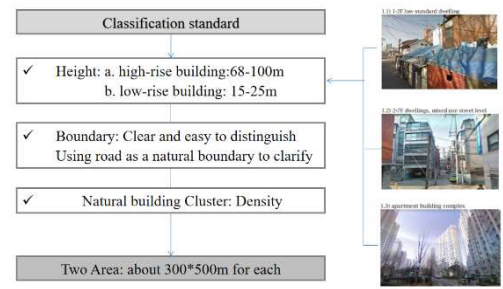

Fig. 4. Classification standard.

Figure 4 illustrates the three principles used to divide the two building blocks. Firstly, building heights were measured using data collected from Naver Map. High-rise buildings were defined as those ranging from 68-100m, while low-rise buildings fell within the 15-25m range [12]. These height ranges served as the overall criteria for classifying the building blocks. Secondly, clear boundaries were established for each building block. This was achieved by analyzing the surrounding road network and using it as a reference, as shown in Figure 5. Thirdly, it was observed that two areas within the site had already formed distinct building blocks, as shown in Figure 6. As a result, two specific building blocks were selected as the research areas for observing the outdoor temperature of high-rise (the left block represented by blue) and low-rise residential buildings (the right block represented by yellow).

The streetscape fabric of the two urban residential zones, along with their related spatial pattern indexes, are outlined below:

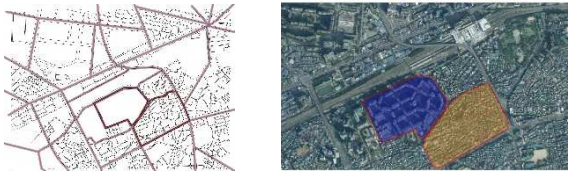

Fig. 5.  Road-network .        Fig. 6.  High-rise and low-rise building.

2) *3D Modeling process:* Through meticulous on-site investigations, we obtained precise specifications for the smallest single building in the selected area. Using a 2D road network map as a foundation, we incorporated green plants along the roadside to ensure the model reflected the actual conditions on the ground, as illustrated in Figure 7. The heights of the buildings in the model were determined based on pre-surveyed data, ranging from 15m to 100m, as depicted in Figure 8. Once the specifications were finalized, we proceeded to model the area with great precision using the Envi-met software, to simulate and analyze the microclimate characteristics of the area in a highly accurate and detailed manner .

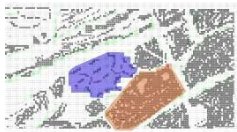 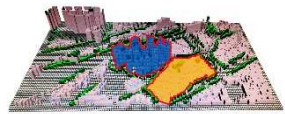

Fig. 7. View of 2D Model.        Fig. 8. View of  3D Model.

3) *Data selection and input:* To establish a stable temperature trend in line with our research objectives, we carefully identified a timeframe characterized by consistent temperature patterns. By choosing a period marked by temperature stability, our aim was to minimize potential confounding factors and enhance the validity of our findings. In ENVI-met, two types of data are required for model calculations. Firstly, the basic data includes time duration and fundamental climate parameters such as temperature, humidity, wind speed, and pressure. We obtained these data from the Korea Meteorological Administration's online public information and inputted them into the software accordingly. Secondly, specific values like relative humidity at 2m or other variables were calculated by referencing the ENVI-met Manual Contents and interpreting the data, as shown in Figure 9.

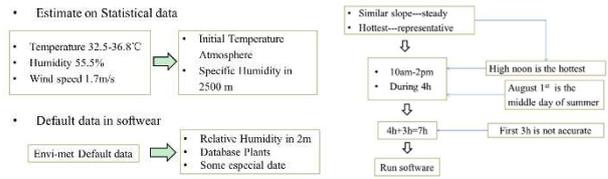

Fig. 9. Data typologies.      Fig. 10. Simulation process in ENVI-met.

The calculation time was also considered. Based on our analysis of the impact of open space size, location, dispersion, and relationships in different residential areas, we found that the microclimate phenomena were primarily concentrated during the daytime. Therefore, our target time period was set from 10 am to 2 pm, with a broader time range from 7 am to 2 pm to account for any inaccuracies in the initial calculation results. With the calculation time and modeling process determined, we initiated the simulation process in ENVI-met, as can be seen in Figure 10.

## IV. RESULTS AND ANALYSIS

### A. Observation of Overall Simulation Results

1) *High-rise Building Block Has Higher Temperature:* By examining the temperature profile at one o'clock in the afternoon (Figure 11), it becomes evident that the high-rise buildings consistently exhibit higher temperatures compared to the low-rise buildings. This temperature difference persists throughout the day.

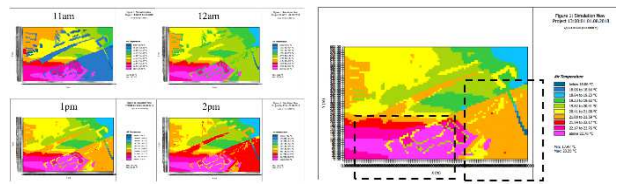

Fig. 11. Simulation result from 11 a.m. to 2 p.m.(left) and  1 p.m.(right).

2) *Open Space Helps to Relieve High Block Temperatures :* In a specific area with 15m buildings and a lower floor area ratio (Figure 12), a significant proportion of open space is present. The temperature profile demonstrates that this area generally experiences lower temperatures, indicating that the presence of open spaces contributes to temperature relief. These green pockets provide shade, promote evapotranspiration, and reduce the heat island effect, all of which contribute to the cooling of the immediate environment.

3) *Air Flow Helps to Lower the Neighborhood Temperature :* The narrow pathway nestled amidst the low-rise buildings (as illustrated in Figure 12) consistently maintains significantly lower temperatures compared to its surrounding areas. The mechanism at play here involves the unrestricted flow of air, which carries away excess heat and promotes heat dispersion, which underscores the beneficial impact of enhanced air circulation in lowering local temperatures.

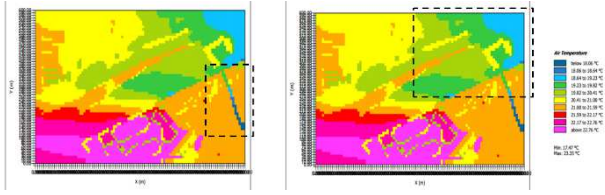

Fig. 12.Simulation result of 1 p.m. of open space(left) and air flow(right).

*B. Analysis and Optimization of Simulation Results*

In this part, the influence of building density, building height, air flow and wind tunnel on the regional microclimate will be analyzed emphatically.

1) *Analysis and Optimization Strategy of Building Density:* Under the condition of similar building heights, higher building density leads to higher temperatures within the block area. Figure 13 shows a notable temperature difference between area 3 and area 2, with area 3 consistently maintaining lower temperatures due to its larger open space. Therefore, when planning building layouts, a balanced consideration of density is crucial to optimize design strategies.

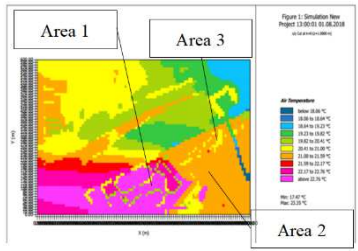

Fig. 13. Simulation result of 1 p.m. of three areas.

2) *Analysis and Optimization Strategy of Building Height:* Higher building heights correspond to higher air temperatures within the block area. As depicted in Figure 15, area 1 consists of high-rise apartment buildings reaching approximately 120 meters in height, while area 2 encompasses ground-floor residential buildings with a height of approximately 15 meters. During the observed periods, area 2 experiences higher temperatures than area 1. Therefore, it is important to strike a reasonable balance in building heights and avoid excessive concentration of high-rise buildings to maintain a balanced outdoor temperature.

3) *Analysis and Optimization Strategy of Air Flow and Wind Tunnel:* High-speed air flow can alleviate high temperatures within the region. Fluid mechanics principles demonstrate that objects obstruct fluid flow, reducing the flow rate. Our findings confirm that high-rise buildings impede air flow in the area, resulting in reduced wind speeds. Previous studies [13] have also supported the impact of high-rise buildings on regional wind speeds. Hence, high-speed air flow proves effective in alleviating regional high temperatures. Additionally, wind tunnels can mitigate local high temperatures. The region indicated as area 1 in Figure 15 represents a narrow road within a densely populated low-rise residential block, acting as a wind tunnel. Due to the swift wind speed within the tunnel, heat dissipates more rapidly, resulting in lower air temperatures compared to the surrounding areas. Therefore, urban morphology considerations should include the incorporation of intervals to create temperature-reducing areas, optimizing design strategies.

In this section, we analyze the influence of building density, building height, air flow, and wind tunnels on regional microclimate. Firstly, we find that higher building density leads to higher temperatures within the block area, emphasizing the need for a balanced consideration of density in design strategies. Secondly, higher building heights correspond to higher air temperatures, highlighting the importance of maintaining a reasonable balance in building heights to prevent excessive concentration of high-rise buildings. Additionally, high-speed air flow and wind tunnels prove effective in alleviating high temperatures. Objects such as high-rise buildings obstruct air flow, reducing wind speeds, while wind tunnels facilitate heat dissipation and result in lower air temperatures. Therefore, optimizing urban design strategies should involve incorporating intervals and temperature-reducing areas to mitigate the adverse effects of building density and height on microclimate.

## V. DISCUSSION

Our research serves as a preliminary investigation into the variations in summer outdoor temperatures and microclimates across diverse residential areas in Yeongdeungpo district. We aim to identify the factors influencing these differences and explore the synergistic effects of various morphological indicators on street spatial texture and microclimate. By conducting simulations and comparative analyses, we aim to offer valuable insights and suggestions to enhance urban climate adaptability through optimized urban spatial morphology in future urbanization developments.

Through our study, we recognize that the design and layout of residential areas play a significant role in shaping the microclimate. The morphological indicators, such as building density, building height, open space, and street layout, collectively impact the local climate. By investigating these factors and their interrelationships, we aim to shed light on the potential strategies and interventions that can be employed to enhance microclimatic conditions. By conducting comprehensive simulations and analyzing the positive microclimate characteristics observed in different residential areas, our research aims to provide practical and actionable recommendations. These suggestions will serve as valuable guidelines for urban planners and developers, helping them optimize urban spatial morphology to improve urban climate adaptability.

Acknowledging the presence of limitations in our research, we are committed to addressing them in future studies. These limitations encompass the restricted research area and limited simulation time. Given time constraints, the simulation time in the ENVI-met software was determined based on representative dates and durations, necessitating further investigation to ascertain the generalizability of our conclusions across different climate scenarios. We are dedicated to ongoing optimization and improvement, with a focus on expanding the research area, improving data collection, and exploring a wider range of climate situations to enhance the reliability and applicability of our findings.

## VI. CONCLUSION

In the context of rapid global urbanization, the field of urban computing holds paramount importance. Understanding the influence of urban microclimates on human well-being is steadily gaining significance. Our research stands as a beacon, offering crucial guidance for sustainable urban design. One pivotal conclusion highlights the need to judiciously avoid the dense clustering of high-rise buildings in future urban designs. Equally crucial is the consideration of an optimal building coverage area and the integration of well-balanced open spaces, especially within central business districts characterized by towering structures. These densely developed urban hubs often generate excessive heat, particularly during summer months, leading to heightened energy consumption for indoor temperature regulation, thus perpetuating the urban heat island effect.

To counteract these challenges, urban designers must adopt a methodical approach to urban morphology and block formation during the design phase. They should proactively anticipate and account for the ensuing microclimate and temperature ramifications over larger areas and future developmental stages. Consequently, the examination of spatial morphology's influence on urban microclimates for optimal urban design emerges as a pivotal research avenue. By embracing more rational and foresighted designs at the planning stage, we can attain sustainable urban environments and nurture design goals that prioritize environmental conscientiousness.

Our research findings carry significant implications for urban development projects and the computation of urban morphology. Specifically, they offer a pivotal approach to mitigating the escalating urban heat island effect and its impact on the microclimate. By integrating the conclusions of our research into their strategic decision-making processes, professionals in the field can actively contribute to the evolution of urban spaces. As stakeholders grapple with the multifaceted challenges posed by contemporary urbanization, our findings underscore the utmost importance of integrating sustainable design strategies. Furthermore, they underscore the imperative of accounting for future climate scenarios and advocating for interdisciplinary collaboration. Through these concerted endeavors, we embark on a transformative journey towards crafting environmentally sustainable cities that prioritize the welfare and comfort of their residents.